\documentclass[%
 aip,
 amsmath,amssymb,
 reprint,%
]{revtex4-1}

\usepackage{graphicx}
\usepackage{dcolumn}
\usepackage{bm}

\usepackage[utf8]{inputenc}
\usepackage[T1]{fontenc}
\usepackage{mathptmx}

\begin{document}

\preprint{AIP/123-QED}

\title[A hybrid model based on deep LSTM for predicting high-dimensional chaotic systems]{A hybrid model based on deep LSTM for predicting high-dimensional chaotic systems}

\author{Youming Lei}
 \email{leiyouming@nwpu.edu.cn.}
 \affiliation{School of mathematics and statistics, Northwestern Polytechnical University, Xi'an 710072, China}
 \affiliation{MIIT Key Laboratory of Dynamics and Control of Complex Systems, Northwestern Polytechnical University, Xi'an 710072, China}
\author{Jian Hu}%
\affiliation{School of mathematics and statistics, Northwestern Polytechnical University, Xi'an 710072, China
}%

\author{Jianpeng Ding}
\affiliation{School of mathematics and statistics, Northwestern Polytechnical University, Xi'an 710072, China
}%

\date{\today}

\begin{abstract}
We propose a hybrid method combining the deep long short-term memory (LSTM) model with the inexact empirical model of dynamical systems to predict high-dimensional chaotic systems. The deep hierarchy is encoded into the LSTM by superimposing multiple recurrent neural network layers and the hybrid model is trained with the Adam optimization algorithm. The statistical results of the Mackey-Glass system and the Kuramoto-Sivashinsky system are obtained under the criteria of root mean square error (RMSE) and anomaly correlation coefficient (ACC) using the singe-layer LSTM, the multi-layer LSTM, and the corresponding hybrid method, respectively. The numerical results show that the proposed method can effectively avoid the rapid divergence of the multi-layer LSTM model when reconstructing chaotic attractors, and demonstrate the feasibility of the combination of deep learning based on the gradient descent method and the empirical model.
\end{abstract}

\maketitle

\begin{quotation}
It is difficult to predict long-term behaviors of chaotic systems only with an inexact empirical mathematical model because of their sensitivity on initial conditions. However, data-driven prediction has made great progress in recent years with the reservoir computing (RC), one of two well-known kinds of recurrent neural network, whose prediction performance is much better than various prediction methods such as the delayed phase space reconstruction. The long short-term memory, as the other kind of recurrent neural network, has the ability to extract the temporal structure of data. Similar to feedforward neural network in structure, it can be easily transformed into a deep one, which is very extensive in practical applications. In this paper, we combine the deep LSTM model with the empirical model of the system to improve the prediction performance significantly with great potential.
\end{quotation}

\section{\label{secI}INTRODUCTION}

The prediction of chaotic systems is one of the most important research fields in recent years. It has been widely used in geological science, meteorological prediction, signal processing and industrial automation. Because of the sensitivity on initial values and the complex fractal structure of chaotic systems, the establishment of a prediction model is a challenging task. At present, this task is mainly accomplished in two ways. One is that over the years, many scientists have established mechanism models such as differential equations by analyzing the internal laws and simplifying main characteristics of natural systems. The other is a data-driven forecasting method, which uses known data sets to build "black box" prediction models. The former requires a lot of expert knowledge. Because of a lack of knowledge on reality systems, the mechanism models have to neglect minor details and are usually approximate. Therefore, it is difficult for chaotic systems to achieve reliable long-term prediction. The advantage of the latter is that it only needs collection of data, regardless of the inherent complex dynamics. But data from reality systems update quickly in the information age nowadays and are difficult to collect on time without errors. There is not an omnipotent data-based model to predict all of the systems since data driven models lack theoretical supports and require tuning parameters in terms of specific problems.

In 2018, Pathak proposed \textbf{a new hybrid method} based on Reserver Computing (RC), which uses the advantages of data and system dynamic structure, and pointed out that this method is also applicable to other machine learning models \cite{pathak2018hybrid}. In fact, both the RC and the LSTM model are derived from general recurrent neural networks. In 2001, Maass and Jaeger proposed a reserve pool calculation for cyclic neural network, respectively, which randomly determines the cyclic hidden unit, so that it could well capture the input history of the sequence in the past, and only learns the output weight, referred to as RC \cite{maass2004fading, jaeger2007echo, jaeger2004harnessing}. It was introduced to predict the trajectories of chaotic systems with good effects \cite{lu2017reservoir, pathak2017using, lu2018attractor, ibanez2018detection, nakai2018machine, antonik2018using, pathak2018model}. The LSTM, on the other hand, is a path that introduces self-circulation to produce gradient continuous flow for a long time \cite{hochreiter1997long, gers2000learning}. As early as 2002, relevant studies have applied the LSTM model to chaotic system prediction, but the effect was not ideal because of the lack of technology at that time\cite{gers2002applying}. Subsequently, numerous studies on data-driven LSTM prediction of chaotic systems have emerged \cite{brunton2019machine, lahmiri2019cryptocurrency}. In 2017, a deep learning network for chaotic system prediction based on noise observation was proposed \cite{yeo2017model}, and the LSTM model was used to filter out noise effectively, and the conditional probability distribution of predicted state variables was realized. In 2018, Vlachas \emph{et al.} applied LSTM to the prediction of high-dimensional chaotic systems, and proved the feasibility of LSTM to predict chaotic systems by comparing it with the GPR method. They also pointed out that LSTM has the potential to mix with other methods to better prediction performance \cite{vlachas2018data}. In 2018, Wan \emph{et al.} combined LSTM with the system's reduced order equation to predict the nonlinear part of systems and the extreme event of complex dynamic equations \cite{wan2018data}. The modeling of time series by RNN can capture the fast and slow changes of the system, and their underlying components are often composed in a layered way.

In 1995, Hihi and Bengio proposed that the deep RNN structure could help extend the ability of RNN to simulate the long-term dependencies of data. They showed that the dependencies of different time scales could be easily and effectively captured by explicitly dividing the hidden units of RNN into groups corresponding to different scales \cite{el1996hierarchical}. Currently, a common method to encode this hierarchy into RNN is to superimpose multiple cyclic neural network layers \cite{graves2013generating, schmidhuber1992learning, graves2013generating, hermans2013training}. There are many variations and improvements of the deep RNN structure model based on the superposition level. For example, in 2015, Chung \emph{et al.} proposed a gated feedforward RNN based on deep LSTM \cite{chung2015gated}. In 2016, Wang \emph{et al.} proposed a regional CNN-LSTM model and applied it to text emotion analysis \cite{wang2016dimensional}. The video caption decoder was simulated by a CNN-LSTM model combined with attention mechanism in 2017 \cite{song2017hierarchical}. To predict chaotic systems more effectively, this work aims to explore the feasibility of combining the deep long short-term memory (LSTM) model, which is based on the gradient descent method, and the empirical model. In practice, we can use all of the resources, data and prerequisite knowledge, and combine them to achieve an optimal prediction result.

The rest of the paper is as follows. In Sec.\ref{secII}, we introduce the structure of the hybrid method based on the multi-layer LSTM model, and describe in detail the training process and prediction process of the method. In Sec.\ref{secIII}, we present three commonly used measures to compare the differences between the single-layer LSTM model, the multi-layer LSTM model, and the hybrid prediction method. In Sec.\ref{secIV}, we demonstrate the feasibility of the hybrid method based on the multi-layer LSTM model in the Mackey-Glass system and in the Kuramoto-Sivashinsky system, respectively. Conclusions are drawn in Sec. V.

\section{\label{secII}HYBRID METHOD}

We consider a dynamical system whose state variable $x(t) \in {R^d}$ is available in the past time. For recurrent neural network (RNN), given input sample $x = ({x_1}, \ldots ,{x_T})$, its hidden vector is $h = ({h_1}, \ldots ,{h_T})$ and the input vector satisfies update equations as follows,
\begin{align}
\label{(1)}
\left \{\aligned
&{h_t} = \varphi ({W_{xh}}{x_t} + {W_{hh}}{h_{t - 1}} + {b_h}),\\
&{y_t} = {W_{hy}}{h_t} + {b_y},
\endaligned
\right.
\end{align}
where $W$ and $b$ represent the weight matrix and the bias vector, and $\varphi $ is the function of the hidden layer.

\begin{figure}
\includegraphics[width=70mm]{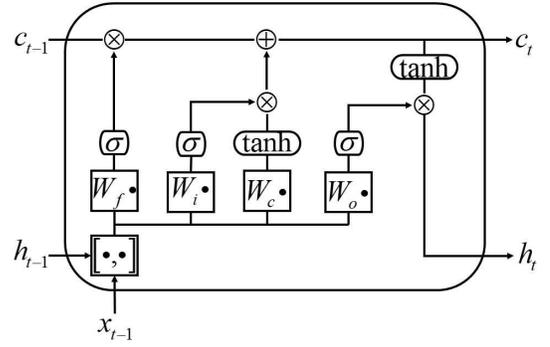}
\caption{\label{fig1} The structure of the "cell" in the LSTM model.}
\end{figure}

In addition to the external circulation of the neural network, LSTM also has an internal "self-loop", shown in Fig.\,\ref{fig1}. For the LSTM model, $\varphi$ is generally realized by the following composite function,
\begin{align}
\label{(2)}
\left \{\aligned
&{i_t} = \sigma ({W_{xi}}{x_t} + {W_{hi}}{h_{t - 1}} + {b_i}),\\
&{f_t} = \sigma ({W_{xf}}{x_t} + {W_{hf}}{h_{t - 1}} + {b_f}),\\
&{o_t} = \sigma ({W_{xo}}{x_t} + {W_{ho}}{h_{t - 1}} + {b_o}),\\
&{c_t} = {f_t}{c_{t - 1}} + {i_t}\tanh ({W_{xc}}{x_t} + {W_{hc}}{h_{t - 1}} + {b_c}),\\
&{h_t} = {o_t}\tanh ({c_t}),
\endaligned
\right.
\end{align}
where $\sigma$ is the sigmoid function, $g_t^f,g_t^i,g_t^o \in {R^{{d_h} \times ({d_h} + {d_i})}}$ represent the forget, input, and output gates respectively, $c,h \in R^{d_h}$ are the "cell" state and the hidden state, ${W_xi},{W_xf},{W_xo},{W_xc} \in {R^{{d_h} \times {d_i}}}$, ${W_hi},{W_hf},{W_ho},{W_xc} \in {R^{{d_h} \times {d_h}}}$ are training weights, and ${b_i},{b_f},{b_o},{b_c} \in {R^{{d_h}}}$ are biases. The LSTM hidden and cell states ($h_t$ and $C_t$ ) are called LSTM states jointly, the dimension of these states is the number of hidden units $d_h$, which controls the ability of cells to learn historical data.

\begin{figure}
\includegraphics[width=70mm]{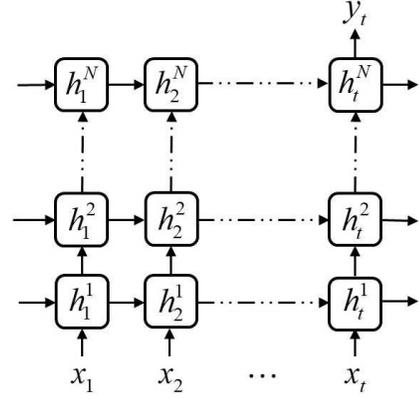}
\caption{\label{fig3} Sample diagram of the deep LSTM model.}
\end{figure}

The process of predicting chaotic system by multi-layer LSTM is shown in Fig.\,\ref{fig3}. Consider a dynamic system, the time series data from  the system $D = \{ {x_{0:N_t}} \} $ are known, where ${x_t} \in {R^{{d_i}}}$ is the system state at time step $t$. The process of fully predicting its state variables by the multi-layer LSTM model is divided into two stages: training and prediction. During the training process, the training set is processed into samples in the corresponding forms of the input and output of the multi-layer LSTM model,
\begin{align}
\label{(6)}
x_t^{train} = \left( \begin{array}{l}
{x_{t + d - 1}}\\
{x_{t + d - 2}}\\
 \vdots \\
{x_t}
\end{array} \right),y_t^{train} = {x_{t + d}},
\end{align}
for $t \in \{ 1,2, \ldots ,{N_{train}} - d + 1\}$. These training samples used to optimize the model parameters to satisfy the mapping relation of ${x_t} \to {y_t}$. The deep architecture learning model can build a higher level interpretation of attractors. For the special structure of LSTM, the deep RNN can be constructed by superimposing several hidden layers easily, where the output sequence of each layer is taken as the input sequence of the next layer, as shown in Fig.\,\ref{fig3}. For the deep RNN constructed by the N-layer LSTM model, the same hidden function is used for all layers in the stack, then the iterative calculation equation of hidden layer sequence $h^n$ of the $n$-th layer
is as follows,
\begin{align}
\label{(3)}
h_t^n = \varphi ({W_{{h^{n - 1}}{h^n}}}h_t^{n - 1} + {W_{{h^n}{h^n}}}h_{t - 1}^n + b_h^n),
\end{align}
where, $n=1,2,...,N$, $t=1,2,...,T$ and $h^0=x$, the output $y_t$ is,
\begin{align}
\label{(4)}
{y_t} = {W_{{h^N}y}}h_t^N.
\end{align}
The loss function of each sample is,
\begin{align}
\label{(7)}
L(x_t^{train},y_t^{train},w) = {\left\| {{F^w}(x_t^{train}) - y_t^{train}} \right\|^2},
\end{align}
where $F^w(x_t^{train})$ represents the output of the model. In the actual training process, to prevent overfitting, regularization terms are added into the error, and the total error function is defined as,
\begin{align}
\label{(8)}
L(D,w) = \frac{1}{S}\sum\limits_{b = 1}^S {L(x_b^{train},y_b^{train},w)} + \lambda \sum\limits_{\omega  \in {W^T}} {{{\left\| \omega  \right\|}^2}},
\end{align}
where $S = {N_{train}} - d + 1$ is the total number of the training samples, and $W^T$ is the set of all trainable parameter metrices. For $N=1$, the model degrades to a single-layer LSTM model.

\begin{figure}
\includegraphics[width=85mm]{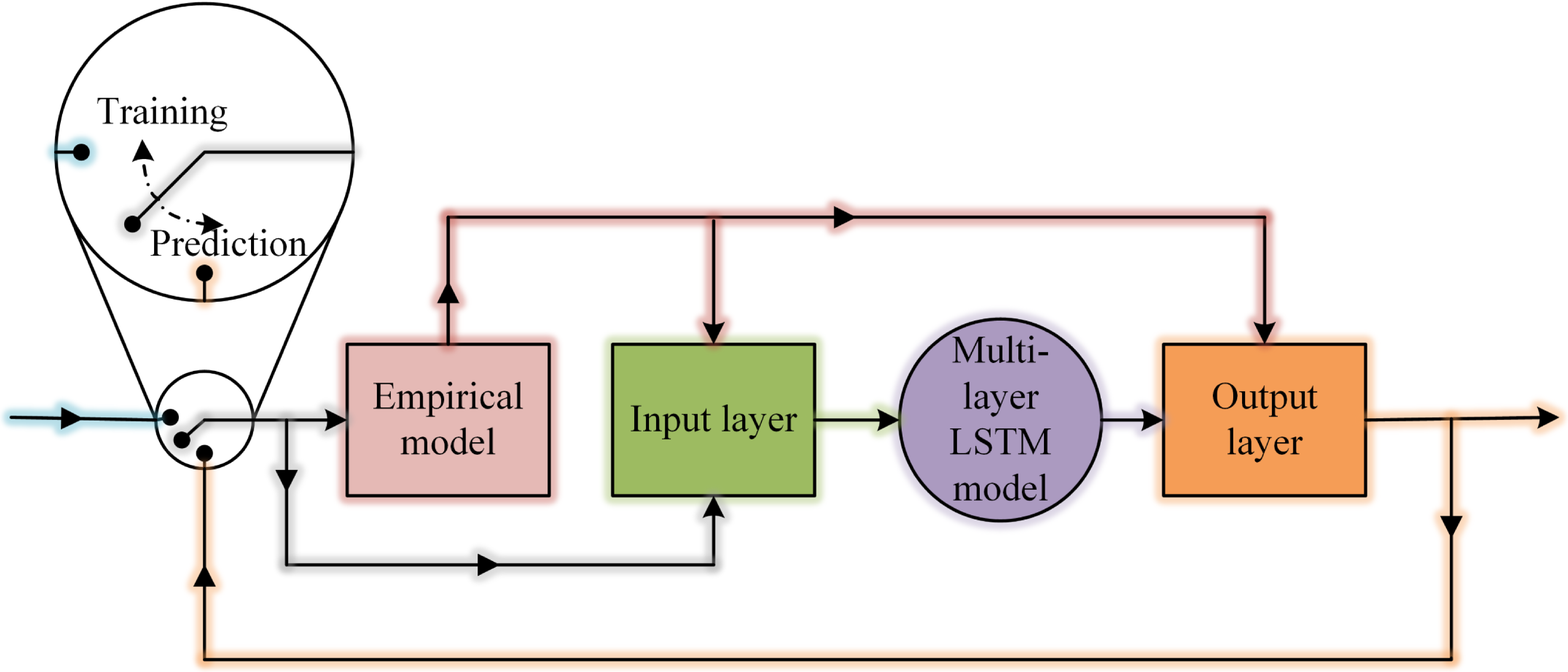}
\caption{\label{fig4} The hybrid method based on multi-layer LSTM, combined the deep LSTM model with the empirical model. }
\end{figure}

At present, deep learning combined with the empirical model was rarely used to predict chaotic systems, and an effective hybrid method based on the reservoir computing was proposed\cite{pathak2018hybrid}. Following it, the structure of combining the multi-layer LSTM model and the empirical model is shown in Fig.\,\ref{fig4}. During the training phase, the training data samples firstly flow into the empirical model to get the empirical data, then flow through the input layer into the multi-layer LSTM model with the empirical data together, and the output of the model and the empirical data finally flow through the output layer to obtain the prediction result. In the prediction phase, the input data is gradually replaced by the prediction results, and the prediction time is iterated until the error reaches the threshold value for the first time.
We assume that the empirical model is an inaccurate differential equation of the dynamical system, whose inaccuracy is represented by parameter mismatch with a small error $\varepsilon $. Although the empirical model only slightly changes system parameters, this can, without loss of generality, generate large errors in an unknown way, which makes long-term prediction impossible only using the mismatch model. In fact, we can consider the parameter mismatch $\varepsilon$ comes from an inevitable error in constructing dynamical equations based on cognitive models. In this case, assume that the state variables in the dynamical system at the previous position are known, and the following state variables can be obtained by integrating the empirical model, so we have
\begin{align}
\label{(11)}
{u^{(\varepsilon)}_{t + \Delta t} } = E[u(t)] \approx u(t + \Delta t).
\end{align}
Then the sample pairs of the hybrid method based on multi-layer LSTM are as follows,
\begin{align}
\label{(12)}
x_t^{train} = \left( \begin{array}{l}
\left[ {{x_{t + d - {\rm{1}}}},E({x_{t + d - {\rm{1}}}})} \right] \\
\left[ {{x_{t + d - 2}},E({x_{t + d - 2}})} \right] \\
 \vdots \\
\left[ {{x_t},E({x_t})} \right]
\end{array} \right),y_t^{train} = {x_{t + d}},
\end{align}
where $t \in \{ 1,2, \ldots ,{N_{train}} - d + 1\}$, $\left[ { \cdot , \cdot } \right] $ represents the splicing of vectors. If $a = {[{a_1},{a_2}, \cdots ,{a_m}]^T}$, $b = {[{b_1},{b_2}, \cdots ,{b_n}]^T}$, $\left[ a,b \right]$ is ${[{a_1},{a_2}, \cdots ,{a_m},{b_1},{b_2}, \cdots ,{b_n}]^T}$. the core part of the model is stacked with multi-layer LSTM as mentioned above, and the output of the model is changed to,
\begin{align}
\label{(13)}
{y_t} = {W_{{h^N}y}}h_t^N + {W_{ey}}E(x(t + d - 1)).
\end{align}

In this work, the same training method is used for different models when comparing the predictive power of different models. In the actual training process, the method of small batch gradient descent is used to solve the optimal parameters. The samples are fed into the model in batches, and the loss function is defined as the mean of the sample loss function for each batch. The training weights are first initialized with Xavier and then iteratively optimized for the network weights. The gradient descent optimizer is used as,
\begin{align}
\label{(9)}
{w^{i + 1}} = {w^i} - \eta {\nabla _w}L(x_t^{train},y_t^{train},{w^i}),
\end{align}
where $\eta $ is the learning rate, $W^i$ is the weight parameter of batch $i$th before optimization, and $W^{i+1}$ is the updated weight parameter. According to different systems, the optimizers used are different. In order to prevent the parameter convergence to a local optimal solution, we use the Adam optimizer, which takes the first and second moment estimation of the gradient into account to accelerate the training speed and effectively avoid the model convergence to the local optimal solution. The updated equations are written as,
\begin{align}
\label{(10)}
\begin{array}{l}
g = {\nabla _w}L(x_t^{train},y_t^{train},{w^i}),\\
m_1^{i + 1} = {\beta _1}m_1^i + (1 - {\beta _1})g,\\
m_2^{i + 1} = {\beta _2}m_2^i + (1 - {\beta _2}){g^2},\\
{{\hat m}_1} = m_1^{i + 1}/(1 - \beta _1^i),\\
{{\hat m}_2} = m_2^{i + 1}/(1 - \beta _2^i),\\
{w_{i + 1}} = {w_i} - \eta {{\hat m}_1}/(\sqrt {{{\hat m}_2}}  + \varepsilon ).
\end{array}
\end{align}

\begin{figure}
\includegraphics[width=90mm]{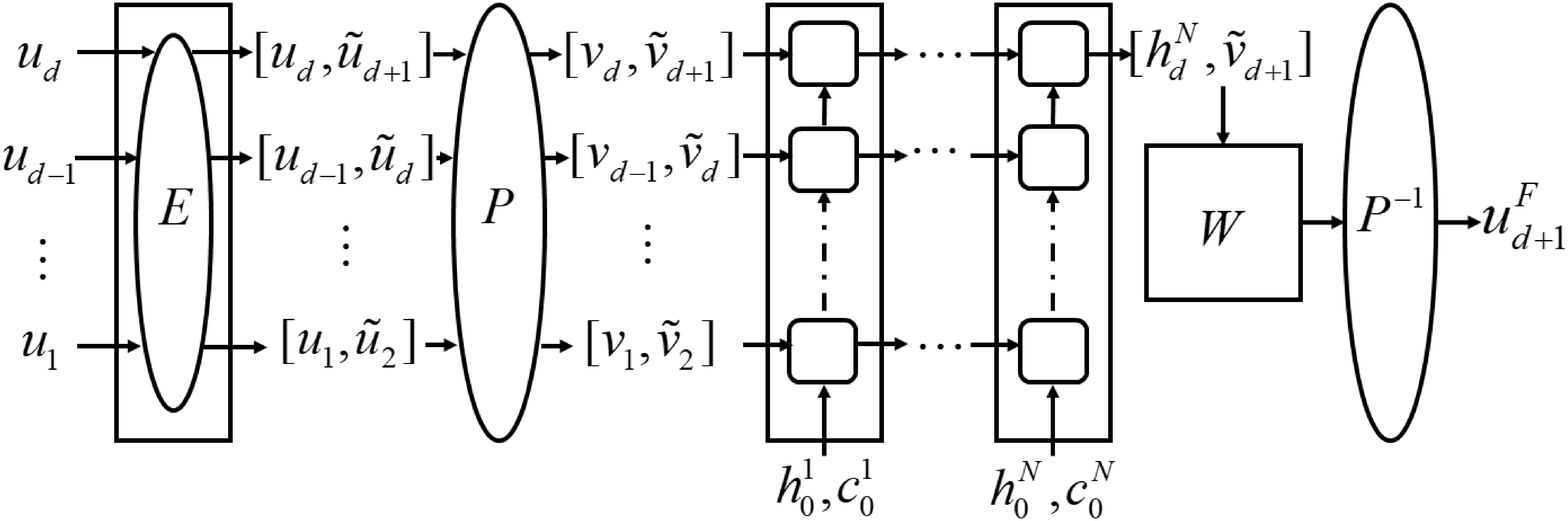}
\caption{\label{fig5} The process flow chart of the hybrid multi-layer LSTM model combined with the empirical model.}
\end{figure}

After each epoch, we update the learning rate by $\eta = \gamma \times \eta$, so that the model remains stable after several training sessions. In order to make the model learn data features more fully, the training samples are normalized during the training process. In the prediction process, the train results of the model are reversed-normalized and taken as the input data in the next step. The complete flow chart is shown in Fig.\,\ref{fig5}. Here, $u_1,u_2,...,u_d$ are the real data of the dynamical system, which are input into the empirical model $E$ to get $\tilde u_1, \tilde u_2,...,\tilde u_d$. Normalize them in $P$, and then input the processed data into the multi-layer LSTM model above. After that, input the output result $\left [ h_d^N, \tilde v_{d+1} \right ]$ into the full connection layer, and thus get the predicted value $u_{d+1}^F$. In the next step, update the input to $u_2,u_3,...,u_{d+1}^F$, and get the corresponding prediction $u_{d+2}^F$. Repeat this progress until we get the final result.

\section{\label{secIII}BENCHMARKS}

We adopt three commonly measures, the normalized error, the root mean square error (RMSE), and the anomaly correlation coefficient (ACC), to compare the prediction performance of different models: singer-layer LSTM, multi-layer LSTM, hybrid singer-layer LSTM, and hybrid multi-layer LSTM. Considering the dimensionless property of the model, we refer to the time unit as the Model Time ($MT$), which is the product of the time step and the number of iteration steps. The normalized error $Exp(t)$ for each independent prediction process is defined as,
\begin{align}
\label{(14)}
Exp(t) = \frac{{\left\| {x(t) - \tilde x(t)} \right\|}}{{{{\left\langle {{{\left\| {x(t)} \right\|}^2}} \right\rangle }^{1/2}}}},
\end{align}
where $x(t)$ is the real data of the chaotic system, and $\tilde x(t)$ is the prediction result of the model. Each method evaluates the effective time by several independent predictive experiments. The effective time $t_v$ is defined as the first time that the normalized error of the prediction model reaches the threshold value $f$, $0<f<1$.
The RMSE is defined as,
\begin{align}
\label{(15)}
{{RMSE}_t} = \sqrt {\frac{1}{V}{{\sum\nolimits_{i = 1}^V {\left\| {x_t^{(i)} - \tilde x_t^{(i)}} \right\|} }^2}} ,
\end{align}
where $V$ refers to the number of different prediction positions selected. In the following, $V=100$. In addition, we used the mean anomaly correlation coefficient (ACC), which is a widely used prediction accuracy index in the meteorological field, at the predicted location to quantify the correlation between the predicted trajectory and the real one. The ACC is defined as,
\begin{align}
\label{(16)}
{{ACC}_t} = \frac{{\sum\limits_{i = 1}^V {{{(x_t^{(i)} - \bar x)}^T}(\tilde x_t^{(i)} - \bar x)} }}{{\sqrt {\sum\limits_{i = 1}^V {{{\left\| {x_t^{(i)} - \bar x} \right\|}^2}} } \sqrt {\sum\limits_{i = 1}^V {{{\left\| {\tilde x_t^{(i)} - \bar x} \right\|}^2}} } }},
\end{align}
where $\bar x$ is the mean value of the training data, and $x_t^{(i)}$ and $\tilde x_t^{(i)}$ represent the predicted and true values of the track at the time of $t$ for the position of $i$-th. The value range of ACC is $[ - 1,1]$, and the maximum value is $1$ while the predicted trajectory and true value change are consistent completely.

\section{\label{secIV}RESULTS}

In this section, different prediction models are applied to two classical chaotic systems: the Mackey-Glass (MG) system and the Kuramoto-Sivashinsky (KS) system.

\subsection{The Mackey-Glass model}

The MG system is a typical time-delayed chaotic system, and the equation is,
\begin{align}
\label{(17)}
\dot x_t = \frac{{\beta {x_{t-\tau} }}}{1 + {x_{t-\tau}^n}} - \gamma x_t,
\end{align}
where $\beta  = 2.0,\gamma  = 1.0,n = 9.65$, ${x_{t-\tau} }$ is the delay term, and $\tau = 2$ represents the time delay. For the MG system, we solve the equation with a second-order Runge-Kutta algorithm with a time step $dt=0.1$ up to $T=10000$ steps, and reconstruct the phase space with the data. The embedded dimension is $8$, and the time delay is $0.1$ to obtain the training data set. Based on the real MG system, the empirical model approximately constructs the chaotic attractor by changing $\gamma $ in Eq.\,(\ref{(17)}) to $\gamma (1 + \varepsilon )$. The error variable $\varepsilon$ is the dimensionless difference between the empirical model and the real model, $\varepsilon=0.05$. For the hybrid method based on the multi-layer LSTM, the attractor can be reconstructed by training data and labels. We construct the deep LSTM model with $N=5$, and the label is given as $y_{train} = x_{t+d} - x_{t+d-1}$. The empirical model data could be obtained by,
\begin{align}
\label{(36)}
&E(x_t) = dt \times \dot x^{(\varepsilon)}_t, \\
&\dot x^{(\varepsilon)}_t = \frac{{\beta {x_{t-\tau} }}}{1 + {x_{t - \tau}^n}} - \gamma(1+\varepsilon) x_t.
\end{align}
These data are reused for $epoch=150$ times, and $batch$ sample pairs are fed for each $epoch$. The hidden dimension of LSTM is $d_h$, and the truncation length of error back propagation is $d$. Parameters of the model are shown in Tabel \ref{tab1},
\begin{table}
\caption{\label{tab1}The parameters for predicting the MG system.}
\begin{ruledtabular}
\begin{tabular}{crcr}
Parameter & Value & Parameter & Value\\
\hline
$batch$   & $20$  & $\eta$    & $0.001$\\
$d$       & $21$  & $\gamma$  & $0.95$\\
$d_h$     & $40$  & $\lambda$ & $5 \times {10^{ - 6}}$ \\
\end{tabular}
\end{ruledtabular}
\end{table}

\begin{figure}
\includegraphics[width=42.5mm]{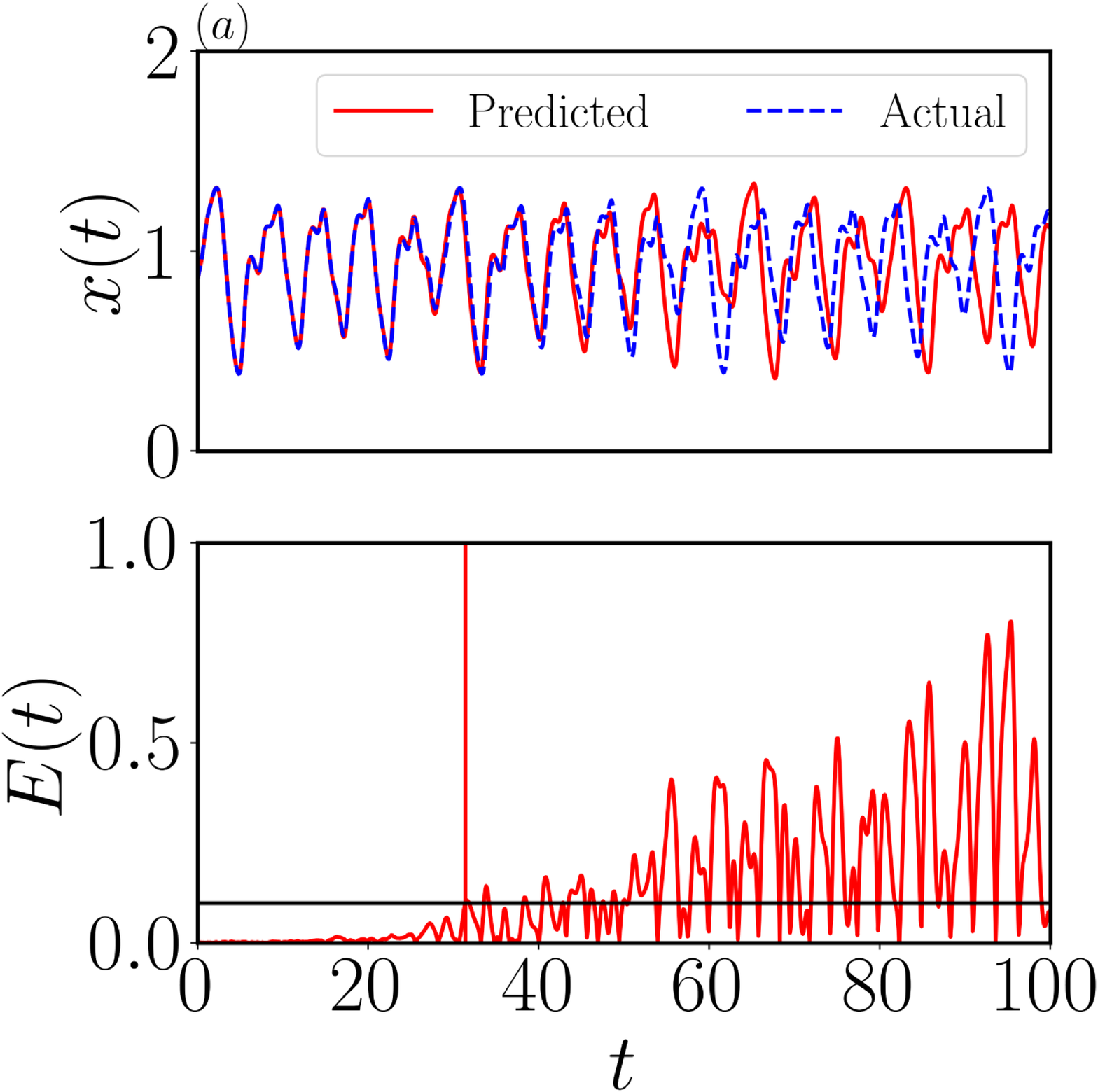}
\includegraphics[width=42.5mm]{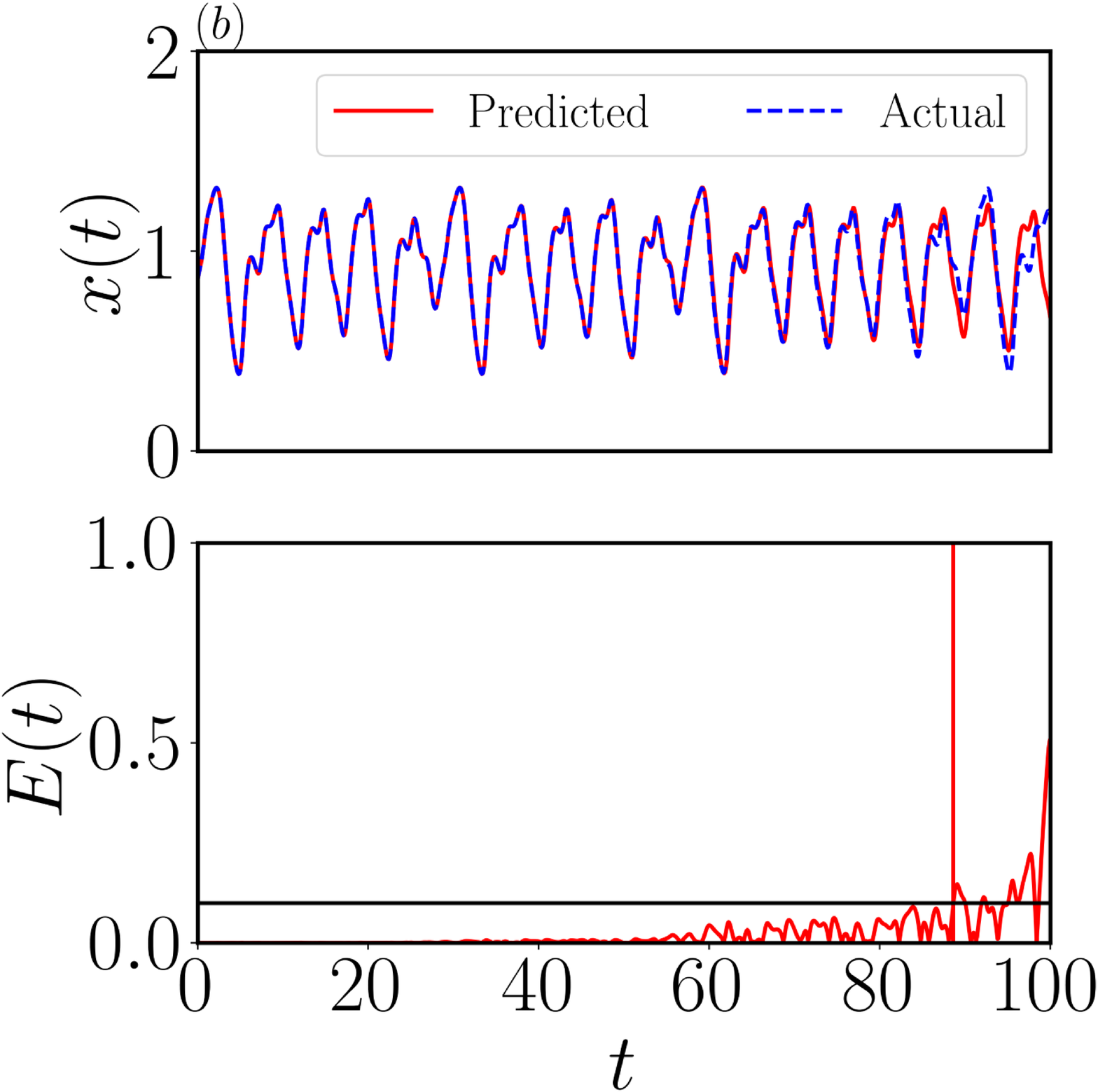}
\caption{\label{fig6} (Color online) Sample diagram of predicting the MG system, (a)the multi-layer LSTM model; (b)the hybrid model based on multi-layer LSTM.}
\end{figure}

\begin{figure}
\includegraphics[width=41.5mm]{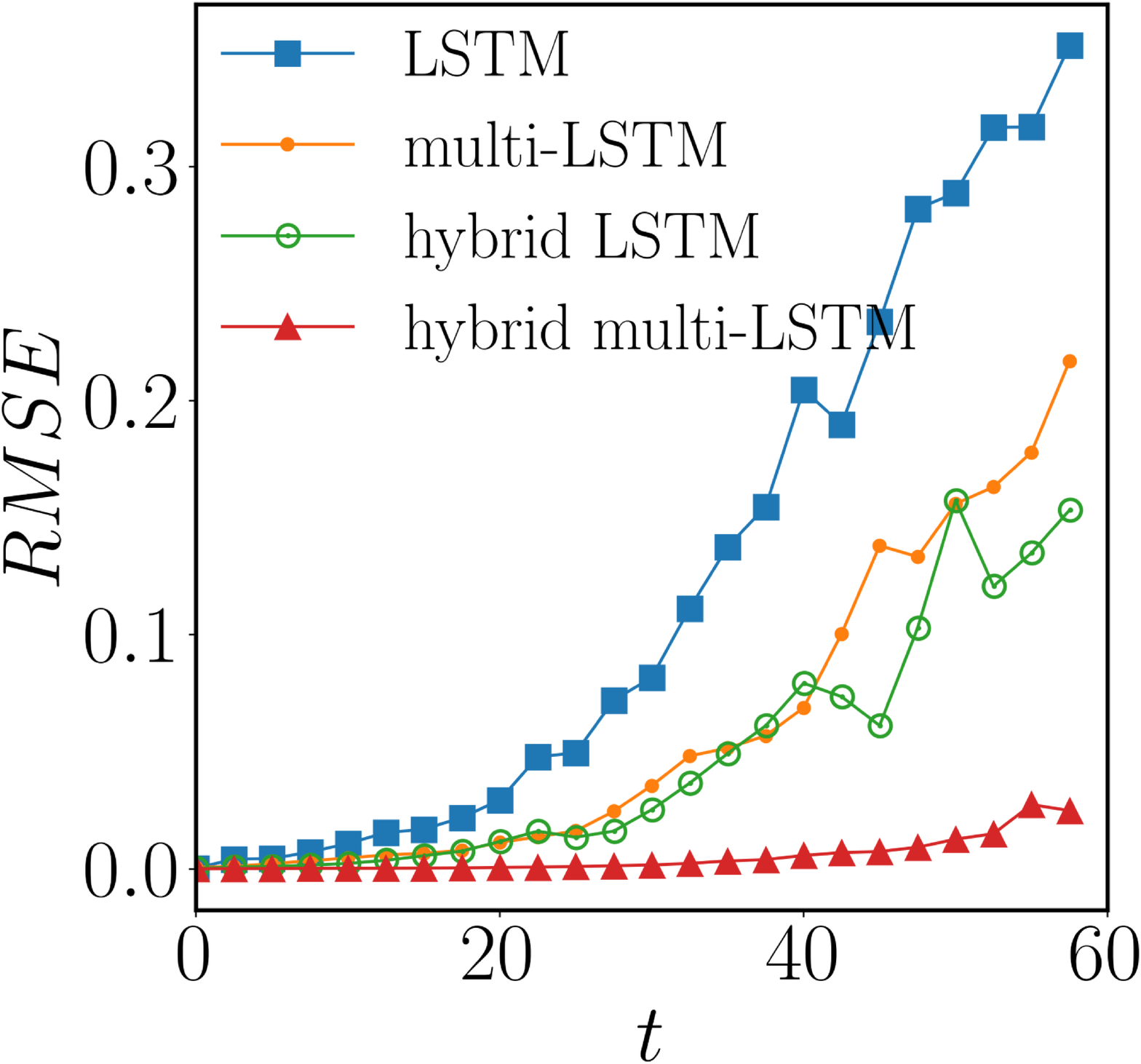}
\includegraphics[width=43mm]{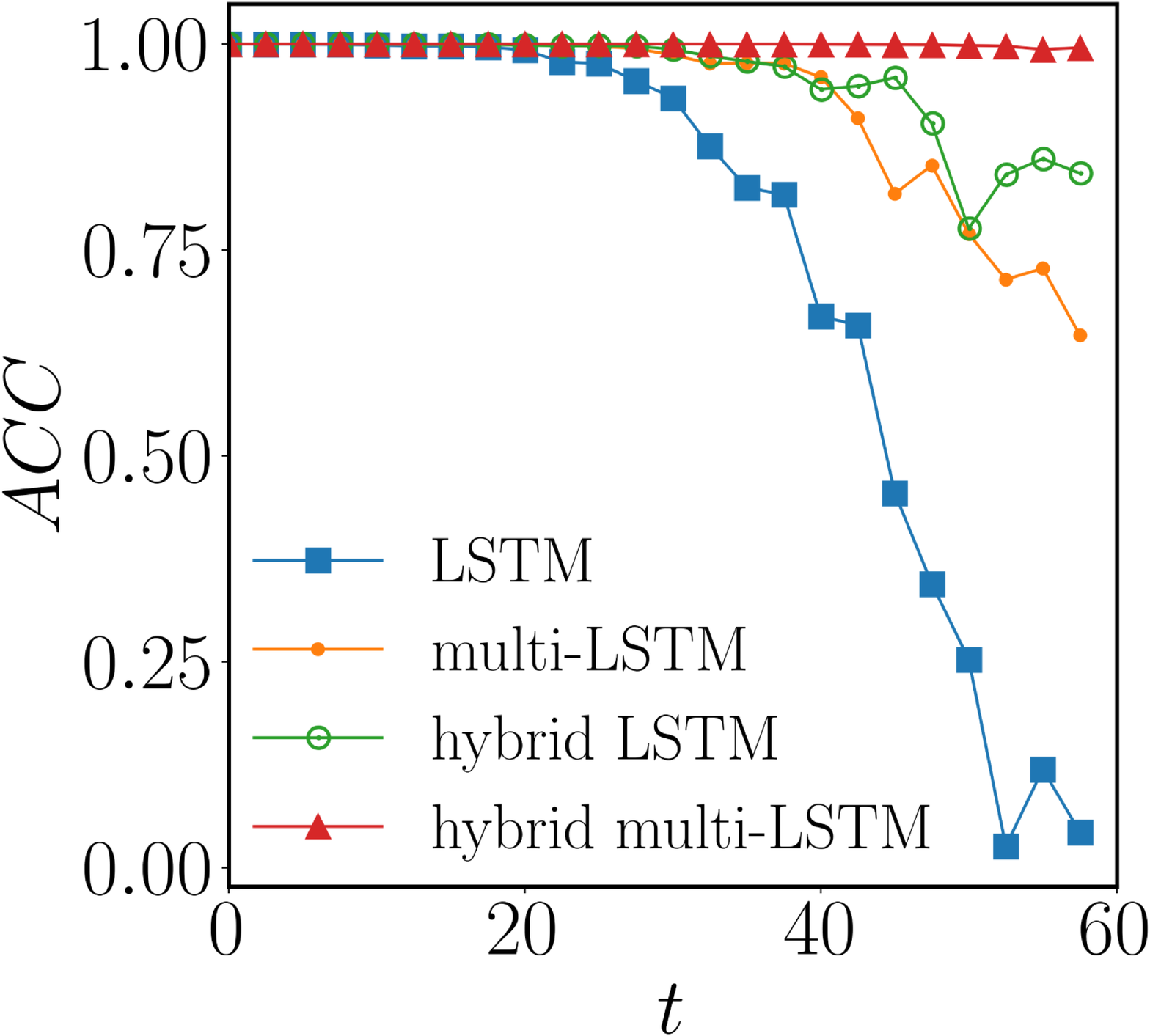}
\caption{\label{fig7} (Color online) The statistical error for predicting the MG system using four models (single-layer LSTM, multi-layer LSTM, hybrid single-layer LSTM and hybrid multi-layer LSTM).}
\end{figure}

Figure\,\ref{fig6} shows the prediction result. The error threshold is $f=0.1$, and the blue dotted line is the real data of the MG system. The prediction time of the hybrid method based on multi-layer LSTM reaches $89.95MT$, which is significantly improved compared with the prediction time $33.02MT$ of the multi-layer LSTM model. Further, the statistical error for predicting $100$ different location by four methods above with the same training data is shown in Fig.\,\ref{fig7}, where the abscissa is forecast time, and the ordinate is the mean of the root mean square error (RMSE) and the mean of anomaly correlatoin coefficient (ACC), respectively. Under the evaluation criteria, the prediction performance can be sorted as: single-layer LSTM $<$ multi-layer LSTM $<$ hybrid single-layer LSTM $<$ hybrid multi-layer LSTM.

\subsection{The Kuramoto-Sivashinsky model}

The KS system is widely used in many scientific fields to simulate large-scale chaotic physical systems, which is deduced by Kuramoto originally \cite{kuramoto1976persistent, kuramoto1978diffusion}. The one-dimensional KS system with initial conditions and boundary is given by,
\begin{align}
\label{(18)}
&\frac{{\partial u}}{{\partial t}} =  - v\frac{{{\partial ^4}u}}{{\partial {x^4}}} - \frac{{{\partial ^2}u}}{{\partial {x^2}}} - u\frac{{\partial u}}{{\partial x}}, \\
&u(0,t) = u(L,t) = {\left. {\frac{{\partial u}}{{\partial x}}} \right|_{x = 0}} = {\left. {\frac{{\partial u}}{{\partial x}}} \right|_{x = L}} = 0.
\end{align}
In order to obtain the real data, discretization via a second-order differences scheme yields,
\begin{eqnarray}
\label{(20)}
\frac{{d{u_i}}}{{dt}}=&&- v\frac{{{u_{i - 2}} - 4{u_{i - 1}} + 6{u_i} - 4{u_{i + 1}} + {u_{i + 2}}}}{{\Delta {x^4}}}\nonumber\\
&&- \frac{{{u_{i + 1}} - 2{u_i} + {u_{i - 1}}}}{{\Delta {x^2}}} - \frac{{u_{i + 1}^2 - u_{i - 1}^2}}{{4\Delta x}},
\end{eqnarray}
where $L=35$ is the periodicity length, the grid size has $D=65$ grid points, and the sampling time is $dt=0.25$. The approximate empirical model is
\begin{eqnarray}
\label{(21)}
&&E({u_t})=u_{t+1}^{(\varepsilon)},\\
&&\frac{{\partial u^{(\varepsilon)}}}{{\partial t}} =  - v\frac{{{\partial ^4}u}}{{\partial {x^4}}} - (1+\varepsilon)\frac{{{\partial ^2}u}}{{\partial {x^2}}} - u\frac{{\partial u}}{{\partial x}},
\end{eqnarray}
where $\varepsilon=0.05$. We use the data of $T=25000$ to train the model with $epoch=150$, and the parameters are shown in Table \ref{tab2}.
\begin{table}
\caption{\label{tab2}The parameters for predicting the KS system.}
\begin{ruledtabular}
\begin{tabular}{crcr}
Parameter & Value & Parameter & Value\\
\hline
$batch$   & $100$  & $\eta$    & $0.001$\\
$d$       & $20$  & $\gamma$  & $0.98$\\
$d_h$     & $50$  & $\lambda$ & $5 \times {10^{ - 10}}$ \\
\end{tabular}
\end{ruledtabular}
\end{table}

\begin{figure}
\includegraphics[width=42.5mm]{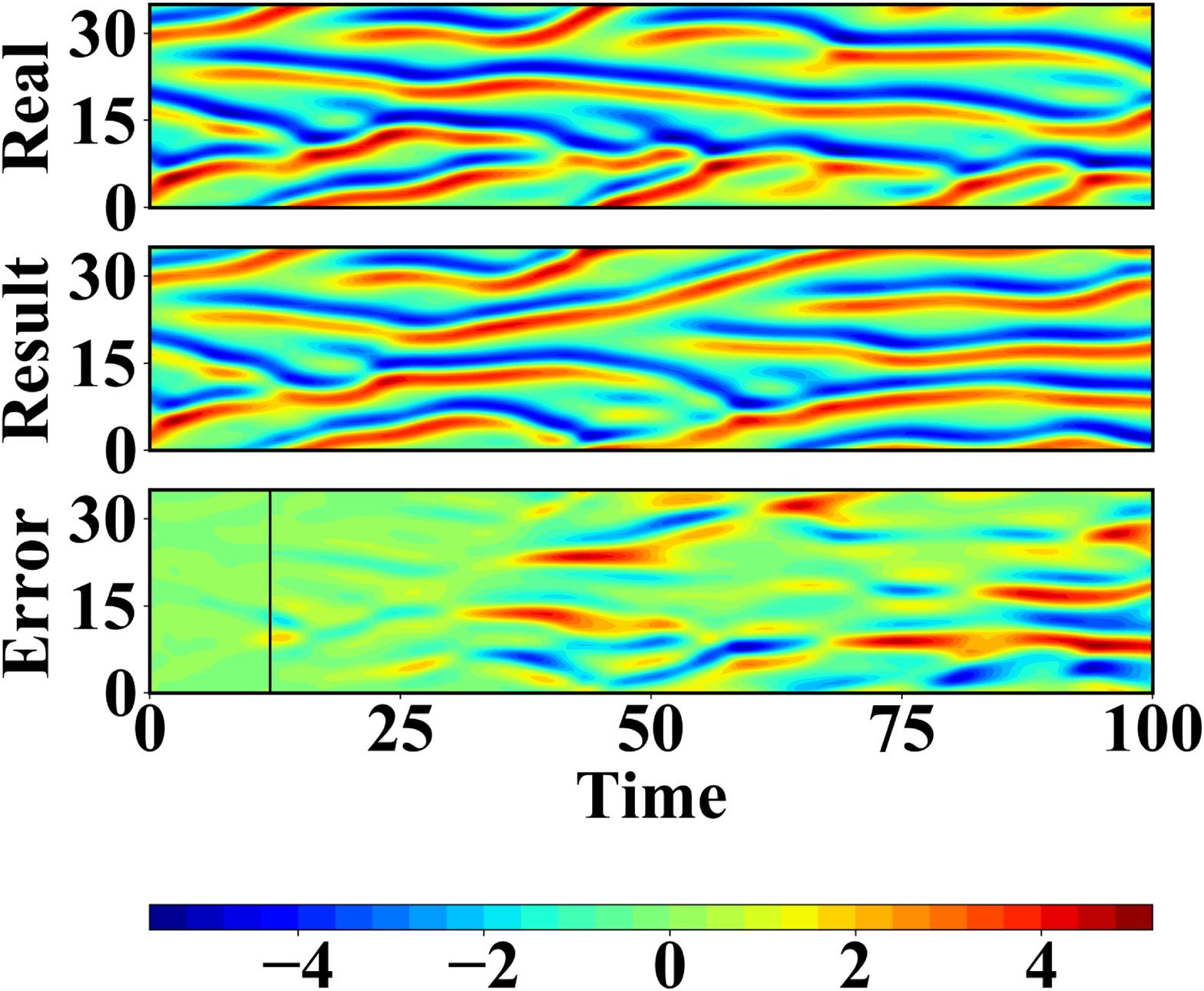}
\includegraphics[width=42.5mm]{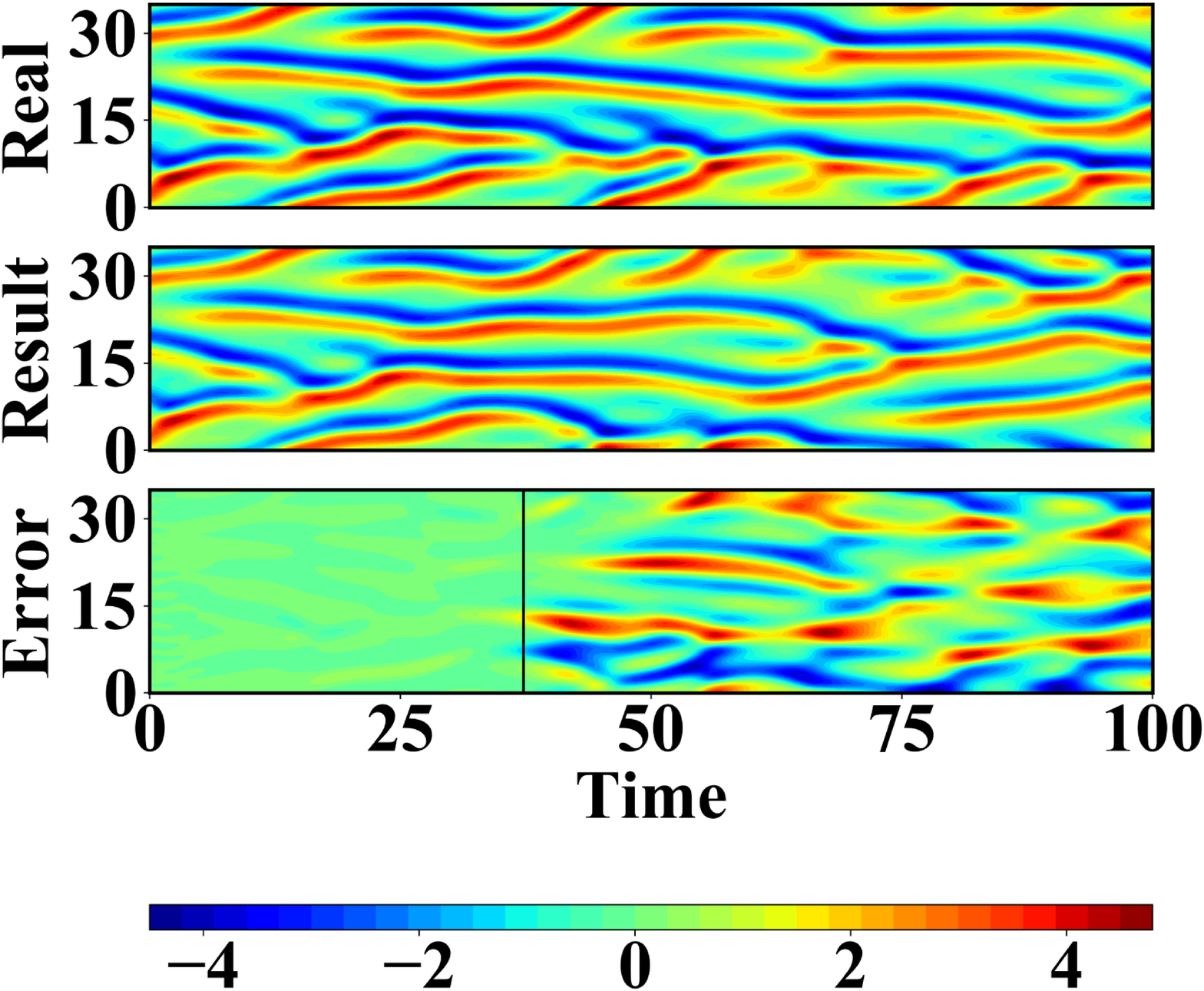}
\caption{\label{fig8} (Color online) Sample diagram of predicting the KS system, (a)the multi-layer LSTM model; (b)the hybrid model based on multi-layer LSTM.}
\end{figure}

\begin{figure}
\includegraphics[width=42mm]{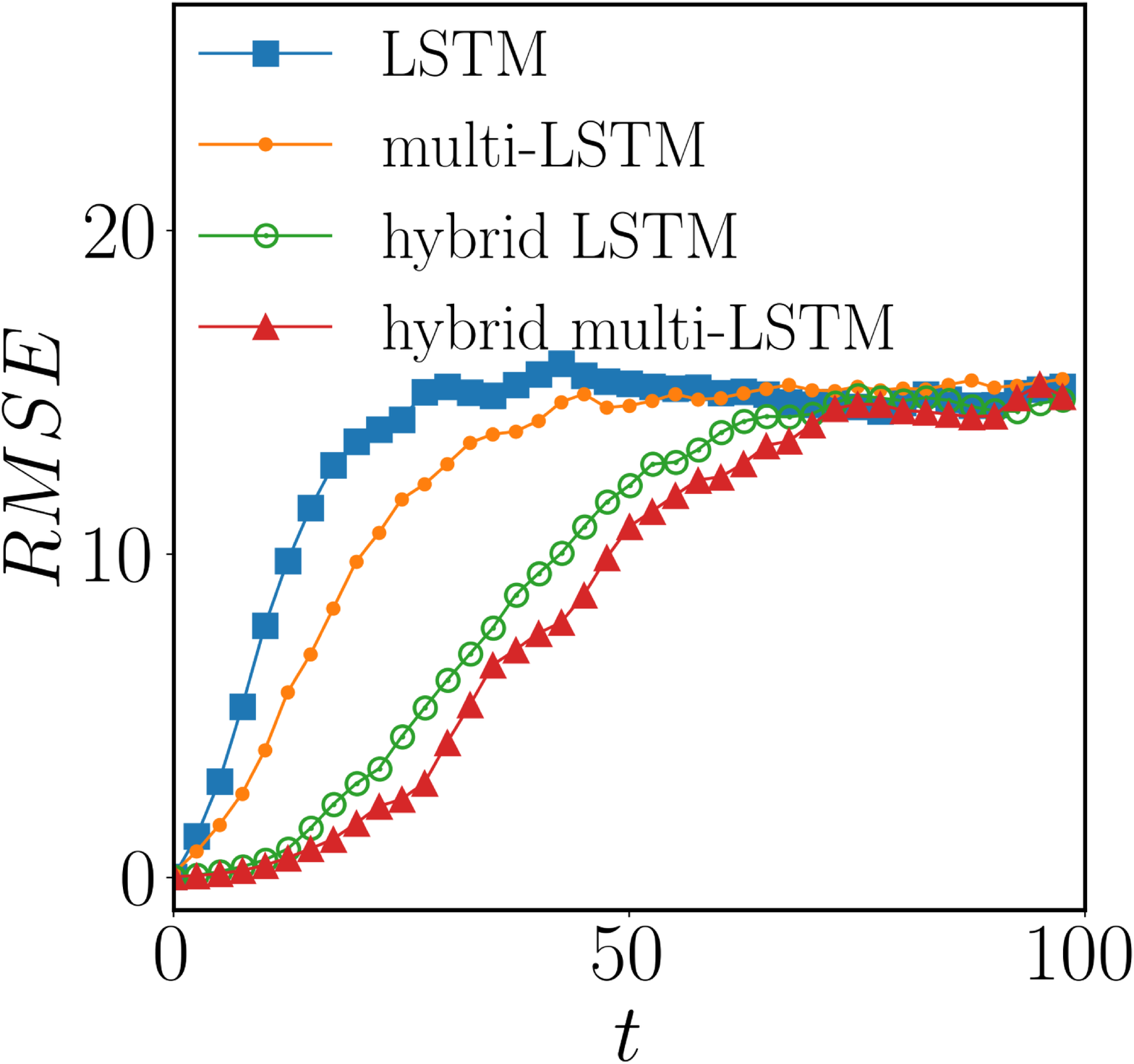}
\includegraphics[width=43mm]{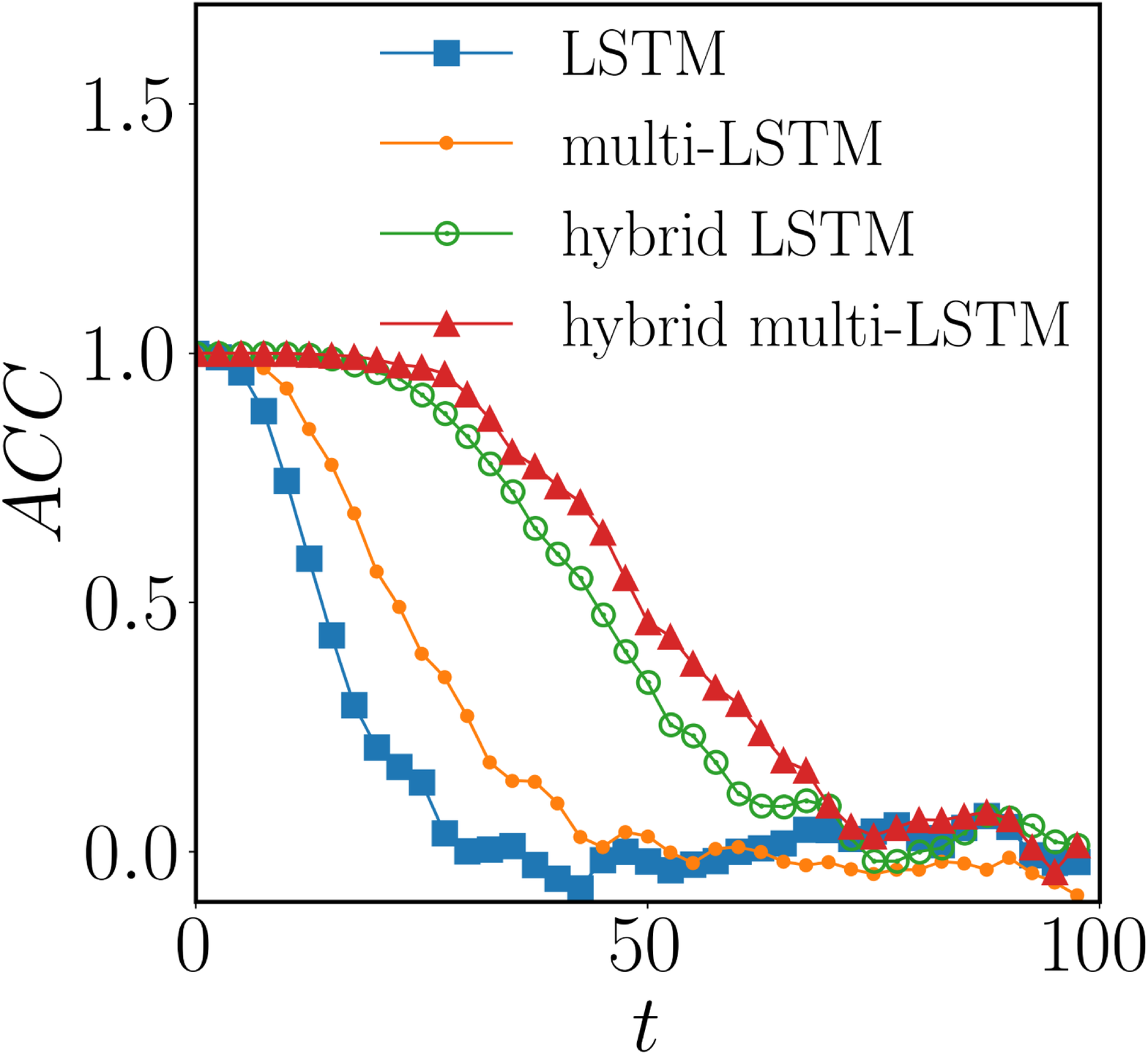}
\caption{\label{fig9} (Color online) The statistical error for predicting the MG system using four models (single-layer LSTM, multi-layer LSTM, hybrid single-layer LSTM and hybrid multi-layer LSTM).}
\end{figure}

Figure\,\ref{fig8} shows the prediction results of the chaotic attractor in the KS system with the multi-layer LSTM model and the hybrid multi-layer LSTM model. In the left figure, given the error threshold of $f=0.4$, the prediction effective time of the multi-layer LSTM model is $12.11MT$, while the prediction time of the hybrid model combined with the empirical model is improved to $38.04MT$. Further, the statistical error for predicting $100$ different locations with four methods above with the same training data is shown in Fig.\,\ref{fig9}, where the abscissa is forecast time, and the ordinate is the mean of the root mean square error (RMSE) and the mean of anomaly correlatoin coefficient (ACC), respectively. Under the evaluation criteria, the prediction performance can be sorted as: single-layer LSTM $<$ multi-layer LSTM $<$ hybrid single-layer LSTM $<$ hybrid multi-layer LSTM. The above numerical results show that for high dimensional chaotic strange attractors, increasing the depth of LSTM model can better the prediction performance to some extent. The hybrid method based on the multi-layer LSTM has significantly improved the prediction capability, which means that the introduction of the empirical model promotes the deep learning method based on the gradient descent method.

\section{\label{V}CONCLUSION}

In this paper, we propose a hybrid method based on the deep LSTM with multi-layers to predict chaos in the MG system and in the KS system, respectively. The LSTM model has a flexible structure and its successful combination with the inexact empirical model provides a feasible way to the prediction of high-dimensional chaos. We can give a heuristic explanation. The data driven LSTM model can capture local characteristics of a chaotic attractor and thus accurately predict the short-term behavior of the system. With time evolution, the divergency of the deep LSTM model from the true trajectory is unavoidable when reconstructing the chaotic attractor. This makes long-term prediction difficult to implement. Introducing the empirical model is just like to tell the hybrid model "climate" of the chaotic attractor so that the deviance between the trajectory and its prediction is corrected to some extent. In this way, combining the deep LSTM model with the empirical model improves the capability of the hybrid model to capture the steady-state behavior of the chaotic attractor while keeping the ability of short-term prediction. \textbf{Note that the prediction performance of the hybrid model based on the deep LSTM is much better than the scheme of RC, while it is similar to the hybrid model combined RC and the empirical model although its training cost is higher than the latter.} Maybe here we have not fully harness the power of the structure of the LSTM. Future work can focus on how to improve the performance of the hybrid model based on a theoretical understanding of the advantage of the LSTM.

\section{\label{VI}Acknowledgement}
This work was supported by the National Natural Science Foundation of China (Grant No. 11672231) and the NSF of Shaanxi Province (Grant No. 2016JM1010).

\nocite{*}
\bibliography{20200114}

\end{document}